# Coriolis force, geometric phase, and spin-electric coupling in semiconductors.

Yuri A. Serebrennikov

Qubit Technology Center

2152 Merokee Dr., Merrick, NY 11566

We consider the response of an effective spin of a charge carrier in semi-conducting systems to an adiabatic rotation of its crystal momentum induced by electric field. We demonstrate that the rotational distortion of Bloch wave functions can be represented by the action of Coriolis pseudo-force that is responsible for torque acting on the orbital momentum of a particle. Mediated by a spin-orbit coupling in the valence band, this perturbation leads to spin rotation that may affect the coherent transport properties of a charge carrier and cause a spin precession in zero magnetic fields. These effects may be also interpreted as a manifestation of, in general, a non-Abelian gauge potential and can be described in purely geometric terms as a consequence of the corresponding holonomy. In the conduction band of wide gap semiconductors, the derived strength of the associated covariant gauge field is proportional to the effective electron $g$-tensor and, hence, is controllable by gate fields or a strain applied to the crystal. The obtained effective spin-Hamiltonians of the carriers in the conduction and light hole bands are homologous to the Rashba Hamiltonian.
72.10.Bg  72.25.Rb  03.65.Vf  03.67.Lx

**I. Introduction.**

In general, the spin and orbital degrees of freedom of a system are coupled, thereby providing the way for an electric field to direct the motion of the spin. In recent years, *all-electric* control over electron spin-precession was demonstrated in 2D semiconductor nanostructures[1]. These results open new opportunities for the design of solid-state quantum computers where the spin rather than the charge of an electron is used for information processing and storage. Clearly, the entirely electric control of the spin may facilitate the integration of spintronics with traditional electronics. Situation becomes especially interesting in the context of the spin-field-transistor proposed by Datta and Das[2] where the current modulation arises from the spin-precession controlled by electric gates.

Besides its importance in technology, interaction of a spin-orbitally (SO) coupled system with an electric field is of considerable interest for our understanding of the fundamental properties of fermionic systems with the time-reversal invariance. The Kramers degeneracy of a Bloch state at zero magnetic fields $B = 0$ makes it a natural candidate for manifestation of non-Abelian gauge potentials and associated holonomies[3]. The latter can be achieved by *adiabatically* driving a *degenerate* system around a closed path in the parameter space, which leads to a nontrivial unitary transformation among the degenerate eigenfunctions of the instantaneous Hamiltonian. It has been shown that the SO-mixed components of a Kramers spinor acquire different geometric phases in an adiabatically revolving external electric field[4,5,6]. The resulting phase shift increasing linearly in time causes a spin-rotation at $B = 0$. The formally equivalent problem has been considered by Zee[7] in the context of nuclear quadrupole resonance[8]. Furthermore, it has been shown that in semiconductors an adiabatic change in the direction of the wave vector $\vec{k}$ of a charge carrier leads to non-trivial gauge potentials that appear in the reciprocal momentum space[9]. The associated covariant gauge field enters the equation of motion for the group velocity of a wave-packet and may affect the coherent transport properties of holes in valence subbands[10][11][12].

In this paper, we extend the results of these studies to the lowest conduction (*C*) band. We show that in the *C*-band of wide gap III-V semiconductors, the strength of the gauge field is proportional to the effective electron *g*-tensor and, hence, is controllable by gate fields or a strain applied to the crystal. In strongly SO coupled valence subbands we recover the results of Refs.[10 - 12], however, the way we derive them here is new. Our procedure illuminates the similarity between the spin-electric coupling in semiconductors and spin-rotation interaction in molecular systems. We demonstrate that the former stems from a rotational distortion of Bloch wave functions, which in the rotating frame of reference is represented by the action of Coriolis pseudo-force and may be also interpreted as a manifestation of generally non-Abelian gauge potential in the momentum space. The geometric approach does not require knowledge of explicit wave functions and is convenient for approximate calculations. Finally, we examine the case that resembles the setup of the Datta and Das spin-field-transistor. We show that in the static uniform electric field applied to the crystal the derived effective spin-Hamiltonians of carriers in the conduction and light-hole bands are homologous to the well-known Rashba Hamiltonian[13].

## II. General Consideration.

In the uniform electric field, $\vec{E}$, each carrier gains the extra momentum $\hbar \delta \vec{k} = q\vec{E}\delta t$, where $q$ is the charge of a particle, and may change its direction in space (in the following $\hbar = 1$). Let the unit vector $\hat{n}^{(L)}$ denote the instantaneous axis of the $\vec{k}$-rotation in the inertial lab (*L*) frame and $\phi$ be the angle between $\vec{k}$ and $\vec{k}' = \vec{k} + \delta\vec{k}$. Geometrically the infinitesimal change in the direction of $\vec{k}$ can be described as $\delta\vec{k}^{(L)} = \delta\phi[\hat{n}^{(L)} \times \vec{k}^{(L)}]$. Hence, if the axis $\hat{n}^{(L)}$ is at the right angle to the plane of the $\vec{k}$ rotation, the instantaneous angular velocity $\vec{\omega}^{(L)} = (d\phi/dt)\hat{n}^{(L)}$ of the $\vec{k} \to \vec{k}'$ rotation

$$\vec{\omega}^{(L)} = [\vec{k}^{(L)} \times \frac{d\vec{k}^{(L)}}{dt}]/k^2 = (q/k^2)\vec{k}^{(L)} \times \vec{E}^{(L)}. \tag{1}$$

Our task is to determine the response of spin-orbitally mixed components of a Kramers-degenerate Bloch state to this rotation. To do this, it is convenient to transform the basis into the moving (*M*) frame of reference that follows the rotation of $\vec{k}$, $\Psi^{(M)}(t) = R^{(L)}(t)\Psi^{(L)}(t)$, where $\Psi$ is the instantaneous eigenvector of the total, *nontruncated* Hamiltonian of the system $H^{(L)}[\vec{k}(t)]$. We assume here that the *M*-frame is rotated with the particle, so that the rotated basis of states is fixed in this frame[14]. If the rotation is uniform, $\vec{\omega}(t) = \vec{\omega}$, the operator $R^{(L)} = \exp[i\vec{\omega}^{(L)}\vec{J}^{(L)}t]$ maps the *L*-frame into the actual orientation of the *M*-frame at time *t*. In our gauge convention, the quantization axis of the system is chosen along the direction of $\hat{k}^{(M)} = \vec{k}^{(M)}/|\vec{k}|$ and corresponds to the $z_M$ axis of the rotating frame. In the presence of SO coupling the Bloch function is

not factorizable into the orbital and spin parts, hence, the total angular momentum, $\vec{J} = \vec{L} + \vec{S}$, is a generator of the rotation $R^{(L)}$. It is easy to see that in the rotating frame[15]

$$\tilde{H}^{(M)}(t) = R^{(L)} H^{(L)}[\vec{k}(t)] R^{(L)+} - i R^{(L)} \dot{R}^{(L)+} = H^{(M)}[|\vec{k}(t)|] - \vec{\omega}^{(M)} \vec{J}^{(M)}, \quad (2)$$

where $\omega_{x_M} dt = -\sin\theta\, d\varphi$, $\omega_{y_M} dt = d\theta$, $\omega_{z_M} dt = \cos\theta\, d\varphi$. The angles, $\theta$ and $\varphi$, specify the orientation of $\hat{k}^{(L)}(\theta, \varphi)$ referred to the $L$-frame at the instance $t$ and provide a different parameterization of the rotation, $R^{(L)} = \exp(i\theta J_{Y_L}) \exp(i\varphi J_{Z_L})$, which yields

$$\tilde{H}^{(M)}(t) = H^{(M)}[|\vec{k}(t)|] - \vec{A}^{(M)} \frac{d\vec{k}^{(M)}}{dt}, \quad (3)$$

whereas,

$$\vec{A}^{(M)} = i R^{(L)}(\hat{k}) \vec{\nabla}_k^{(L)} R^{(L)+}(\hat{k}). \quad (4)$$

Thus, in the $M$-frame the pure gauge potential $\vec{A}^{(M)}$ can be expressed as

$$\vec{A}^{(M)} \frac{d\vec{k}^{(M)}}{dt} = \vec{\omega}^{(M)} \vec{J}^{(M)} \quad (5)$$

or, equivalently, as

$$\vec{A}^{(M)} = -[\vec{k}^{(M)} \times \vec{J}^{(M)}]/k^2. \quad (6)$$

The term $\vec{\omega}^{(M)} \cdot \vec{J}^{(M)}$ represents the combined effect of the Coriolis and centrifugal terms, $-\vec{\omega}^{(M)} \vec{L}^{(M)}$, and spin-rotation interaction, $-\vec{\omega}^{(M)} \vec{S}^{(M)}$, as seen by an observer in the noninertial $M$-frame[16]. Up to this point, the transformations are exact, $\vec{A}$ is a pure gauge potential and there is no covariant gauge field associated with it. The next step entails adiabaticity of the wave vector's rotation and leads to nontrivial gauge potentials, covariant gauge fields, and related holonomies.

To clarify the underlining physics, let us ignore, for a moment, the spin degree of freedom and consider the single particle Hamiltonian $\widetilde{H}^{(M)} = k^2/2m_0 - \vec{\omega}^{(M)} \vec{L}^{(M)}$, where $m_0$ is the bare electron mass. Taking into account that $\vec{L}^{(M)} = [\vec{r}^{(M)} \times \vec{k}^{(M)}]$, we have

$$\widetilde{H}^{(M)} = (\vec{k}^{(M)} - m_0 [\vec{\omega}^{(M)} \times \vec{r}^{(M)}])^2/2m_0 - [\vec{\omega}^{(M)} \times \vec{r}^{(M)}]^2 m_0/2, \quad (7)$$

which corresponds to the classical Hamiltonian of a particle in an uniformly rotating frame[17]. The last term can be identified as the centrifugal potential energy of a particle, whereas the first one reflects the gauge dependence of the canonical momentum. The assumed slowness of $\vec{k}$ rotation allows to ignore terms $\sim \omega^2$. Consequently, upon calculating the commutators of $\widetilde{H}^{(M)}$ with $\vec{k}^{(M)}$ and $\vec{r}^{(M)}$ one finds

$$d\vec{r}^{(M)}/dt = \vec{k}^{(M)}/m_0 - [\vec{\omega}^{(M)} \times \vec{r}^{(M)}], \quad d\vec{k}^{(M)}/dt = -[\vec{\omega}^{(M)} \times \vec{k}^{(M)}]. \quad (8)$$

Note that in this approximation the results of dynamic and pure geometric considerations are essentially the same. The minus sign of the vector products in Eq.(8) reflects the point of view of a rotating observer. In a classical description, the term $[\vec{k}^{(M)} \times \vec{\omega}^{(M)}]$ represents Coriolis pseudo-force acting on a particle in the rotating frame. It is easy to see that this force causes torque $d\vec{L}^{(M)}/dt = i[\widetilde{H}^{(M)}, \vec{L}^{(M)}] = -\vec{r}^{(M)} \times [\vec{k}^{(M)} \times \vec{\omega}^{(M)}]$ acting on the orbital momentum of a particle. In the Schrödinger representation this perturbation reflects a rotational distortion of the wave functions[15] with orbital momentum $L \neq 0$.

In the presence of spin, the Bloch states in crystals with inversion symmetry are doubly degenerate at $B = 0$. Within the adiabatic approximation, i.e., when the interband distances are much larger than $|\vec{\omega}|$ at any instantaneous orientation of $\vec{k}$ we may define a projector $P_B^{(M)}$ onto the complex 2D Hilbert space spanned by the two-component

Bloch spinor functions $\Psi_B^{(M)}(t) = P_B^{(M)} \Psi^{(M)}(t)$, where $B = C$, $LH$, and $HH$ indicates the relevant band. The indices $LH$ and $HH$ denote the light and heavy holes, respectively. The appropriate adiabatic basis, fixed in the rotating frame, is defined by the eigenvectors of the Hamiltonian, $H^{(M)}$, which may correspond to, e.g., $M$-frame 8x8 Kane Hamiltonian. The projection of the Eq.(2) onto this basis yields the following Schrödinger-type equation

$$i \dot{\Psi}_B^{(M)}(t) = \widetilde{H}_B^{(M)} \Psi_B^{(M)}(t), \qquad (9)$$

$$\widetilde{H}_B^{(M)} = k^2/2m_B - \vec{\omega}^{(M)} \vec{\tilde{\gamma}}_B^{(M)} \vec{\sigma}^{(M)}/2 = k^2/2m_B - \vec{A}_B^{(M)} \frac{d\vec{k}^{(M)}}{dt}, \qquad (10)$$

where $\vec{A}_B^{(M)} = \hat{e}_\theta (\vec{A}_B^{(M)})_\theta / k + \hat{e}_\varphi (\vec{A}_B^{(M)})_\varphi / (k \sin\theta)$ is the momentum space Wilczek-Zee gauge potential and

$$(\vec{A}_B^{(M)})_\theta = \gamma_{\perp,B} \sigma_{Y_M}/2, \quad (\vec{A}_B^{(M)})_\varphi = (\gamma_{\parallel,B} \cos\theta \, \sigma_{Z_M} - \gamma_{\perp,B} \sin\theta \, \sigma_{X_M})/2. \qquad (11)$$

Here $\vec{\sigma}$ is the vector of Pauli matrices, $m_B$ is the effective mass of a carrier in the $B$-band, which is assumed isotropic. The "tensor" $\vec{\tilde{\gamma}}_B$ is defined by the expression[5][18]

$$\vec{\tilde{\gamma}}_B^{(M)} \vec{\sigma}^{(M)}/2 := P_B^{(M)} \vec{J}^{(M)} P_B^{(M)}. \qquad (12)$$

The symmetry of the problem suggests that $(\vec{\tilde{\gamma}}_B)_{x_M x_M} = (\vec{\tilde{\gamma}}_B)_{y_M y_M} = \gamma_{\perp,B}$, $(\vec{\tilde{\gamma}}_B)_{z_M z_M} = \gamma_{\parallel,B}$. Obviously, $\vec{\tilde{\gamma}}_B$ is not a true tensor, since it does not transform covariantly under gauge transformations. The Schrödinger-type equation (9) and expressions (10) - (12) depend on a choice of gauge that specifies the reference orientation, i.e. the orientation in which the $M$-frame coincides with some space-fixed frame. At the moment $t = 0$, this orientation may always be chosen (*locally* in the $\vec{k}$-space) such that $\vec{\tilde{\gamma}}_B^{(M)}$ is diagonal in the helicity basis, and that the main axis $z_M$ of this "tensor" represents the quantization axis of the pseudospin operator $\vec{\sigma}/2$.

The term $\vec{\omega}^{(M)} \vec{\gamma}_B^{(M)} \vec{\sigma}^{(M)}/2$ in our Hamiltonian Eq.(10) can be viewed as a generic Zeeman Hamiltonian of a spin-1/2 particle in a *fictitious* magnetic field $\vec{\omega}^{(M)} \vec{\gamma}_B^{(M)}$ and will split the components of a Kramers doublet even in the absence of an external magnetic field. Noticeably, this term is equivalent to the *M*-frame Hamiltonian of spin-rotation interaction in molecular systems[5] and can be interpreted as a manifestation of generally non-Abelian gauge potential in the momentum space. The covariant field associated with the gauge potential $\vec{A}_B^{(M)}$ can be readily derived from Eq.(11)

$$\vec{F}_B^{(M)} = \vec{\nabla}_k^{(M)} \times \vec{A}_B^{(M)} - i[\vec{A}_B^{(M)}, \vec{A}_B^{(M)}] =$$
$$[(\gamma_{\perp,B}^2 - \gamma_{\|,B})\sigma_{Z_M} + \gamma_{\perp,B}(\gamma_{\|,B} - 1) ctg\theta \, \sigma_{X_M}](\hat{k}^{(M)}/2k^2) = \quad (13)$$
$$(\gamma_{\perp,B}^2 - \gamma_{\|,B})(\hat{k}^{(M)}/k^2) \sigma_{Z_M}/2$$

Here we took into account that for any eigenstates of $H^{(M)}$ with the helicity $m = \hat{k}_{Z_M} J_{Z_M} = \pm 1/2, \gamma_{\|,B} \equiv 1$; and for $|m| > 1/2, \gamma_{\perp,B} \equiv 0$ (see also sec. III). As seen from Eq.(13), the gauge invariant strength of this field is proportional to the strength of the Dirac monopole at the origin of the momentum space[3], whereas $\gamma_{\perp,B}^2 - \gamma_{\|,B}$ plays the role of a "screening" parameter. It is important to stress here that the field $\vec{F}_B$ enters the equation of motion for the expectation value of the real-space position $<\vec{r}^{(M)}>_B = <\Psi_B^{(M)}|\vec{r}^{(M)}|\Psi_B^{(M)}>$ of the center of the wave packet that represents a charge carrier[11][12][13]

$$<\dot{\vec{r}}^{(M)}>_B = <\vec{k}^{(M)}>_B / m_B + <\vec{F}_B^{(M)}> \times <\dot{\vec{k}}^{(M)}>_B \quad (14)$$

and may affect the coherent motion of electrons and holes in adiabatically isolated bands.

Clearly, the presence of an electric field that has a fixed direction in space cannot break the time-reversal symmetry and split or induce transitions between the components of a Kramers' doublet. Nevertheless, we see that rotation of the crystal momentum induced by the static $\vec{E}$ violates the T-invariance of a system and due to SO coupling leads to the specific form of spin-electric coupling: spin-rotation interaction. In the absence of SO interaction the Bloch wave functions for any $\vec{k}$ are merely products of orbital and spin functions. If the C-band $\Psi_C^{(M)}$ were a pure s-like ($L = 0$) spin doublet, then these functions would span a complete representation of the spinor group *SU(2)*, $\vec{A}_C^{(M)}$ would become the pure gauge with no covariant gauge field associated with it ($\vec{\gamma}_C^{(M)} = \vec{1}$, $F_C^{(M)} = 0$). In this case, spin will be entirely decoupled from the rotation of $\vec{k}$. Moreover, since the Coriolis and SO interactions involve the orbital angular momentum, both would vanish to first order in the C-band were it not for the admixture of p-like ($L = 1$) valence states. Hence, similarly to molecular systems, spin rotation in the C-band entails the "s-p hybridization" of the adiabatic functions $\Psi_C^{(M)}$, which requires that the symmetry of the system is lower than spherical. Fortunately, even in crystals with inversion symmetry translational motion of a carrier always breaks the isotropy of a system and is responsible for the anisotropic part of an *instantaneous* $\vec{k} \cdot \vec{p}$ Hamiltonian, which reflects the coupling between the local effective orbital moment and the lattice momentum of a particle[19]. As a result, the mixed states $\Psi_B^{(M)}$ neither span a complete representation of the *SU(2)* in the C-band nor of the $SU(2) \times SO(3)$ double group in strongly SO coupled valence bands. Consequently, in common semiconductors, spin of a carrier is coupled to the rotation of $\vec{k}$; strength of the field $|\vec{F}_B^{(M)}| \sim |\gamma_{\perp,B}^2 - \gamma_{\parallel,B}|$ associated with non-Abelian gauge potential $\vec{A}_B^{(M)}$ is, in general, not zero.

An electric field enters the problem through the *parametric* dependence of the *L*-frame Hamiltonian on $\hat{k}^{(L)}$. Appearance of the Wilczek-Zee gauge potential in Eq.(10) becomes clear if we take into account that a differential action of $\vec{\omega}^{(M)} \vec{\tilde{\gamma}}_B^{(M)} \vec{\sigma}^{(M)}/2$ is proportional to the angle of rotation, $|\vec{\omega}(t)|dt$, i.e., to the *distance* in the angular space. Correspondingly, $\vec{A}_B^{(M)} d\vec{k}^{(M)}$ provides a pure geometrical mapping between an infinitesimal change in the orientation of $\vec{k}$ in the 3D-Euclidean momentum space and the resultant rotation of $\Psi_B^{(M)}$ in the 2D spinor space. As the system slowly rotates, it adiabatically passes through an infinite sequence of configurations in the angular space that can be parameterized by the angle-axis $\{\phi, \hat{n}^{(M)}\}$ or by the $\{\theta, \varphi\}$ variables. Results do not depend on the actual physical mechanism of this rotation nor they depend on the charge carrier being an electron or a hole. As long as $\vec{k}$-rotation represents an adiabatic perturbation to the system, the evolution of the spinor $\Psi_B^{(M)}$ is a unique function of the curve traversed by the wave vector in the angular space and is independent of the rate of traversal. For finite times, the infinitesimal rotations of a spinor accumulate to a finite rotation, thereby giving rise to transitions between the pair of Kramers-conjugate states. In general, the axis of rotation may change its direction in time, so the elementary rotations of $\vec{k}$ may not commute. Nonetheless, the formalism remains the same. To find the evolution of $\Psi_B^{(M)}$ in this situation one must evaluate the path-ordered integral along the curve traversed by the wave vector in the angular space.

To describe the evolution of the Bloch state in the *local inertial* (reference) frame we have to perform a reverse rotation of the coordinate system compensating for the rotation of the *M*-frame, thereby closing the path in the angular space by the geodesic. This transformation is not associated with a physical change of a state and does not affect the kinetic energy of the carrier. In the 2D projective spinor-space, it is merely a $R_{KD}^{(L)+} = \exp[-i\vec{\omega}^{(L)} \vec{\sigma}^{(L)} t/2]$, which yields

$$i \overset{\bullet}{\Psi}_B^{(L)}(t) = [k^2/2m_B + H_{B,SR}^{(L)}] \Psi_B^{(L)}(t), \qquad (15)$$

$$H_{B,SR}^{(L)} := -\vec{\omega}^{(L)} \Delta \vec{\tilde{\gamma}}_B^{(L)} \vec{\sigma}^{(L)}/2, \qquad (16)$$

where $\Delta \vec{\tilde{\gamma}}_B^{(L)} := \vec{\tilde{\gamma}}_B^{(L)} - \hat{1}$. Remarkably, the *effective* spin-Hamiltonian, Eq.(16), has a familiar form of spin rotation interaction in molecular systems[5]. The original non-truncated multiband Hamiltonian of the problem serves to determine the gauge group and the principal values of $\vec{\tilde{\gamma}}_B$. The underlining physics of the problem is hidden in the definition of $\vec{\tilde{\gamma}}_B, \vec{\omega}$, and the effective mass of a carrier.

**III. Valence and Conduction Bands in III-V Semiconductors.**

Similarly to a fine structure splitting in isolated atoms, SO interaction breaks up the six-fold valence band degeneracy at $\Gamma$-point into multiplets of $J$ ($J = 3/2$, $\Gamma_8$ and $J = 1/2$, $\Gamma_7$), but preserves the isotropy of the system. Anisotropy comes from the translational motion of the hole that shifts the carrier from the center of the zone and, analogous to a crystal field, is responsible for further lifting of the degeneracy of the $\Gamma_8$ states into *HH* and *LH* bands. The *M*-frame four-band Luttinger Hamiltonian[20] can be expressed in the following form $H_{3/2}^{(M)} = (k^2/2m_0)[\gamma_1 + 2\gamma_2(J^2/3 - J_{Z_M}^2)]$, where $J = 3/2$ and the coefficients $\gamma_{1,2}$ are the dimensionless Luttinger parameters. Due to the *T*-invariance of the problem eigenvalues of $H_{3/2}^{(M)}$ have Kramers degeneracy. We note that $H_{3/2}^{(M)}$ is axially symmetric and is diagonal in the $|LS, Jm>$ basis. Hence, $J_{Z_M}$ is conserved and the eigenfunctions of this Hamiltonian can be classified by the helicity $m = \hat{k}_{Z_M} J_{Z_M}$. Bands with $m = \pm 3/2$ correspond to *HH*s; whereas bands with $m = \pm 1/2$ represent *LH*s. Adiabaticity of the $\vec{k}$-rotation means that $\omega$, is much smaller than the energy separation between the *HH* and *LH* bands determined by the anisotropic part of the *instantaneous* Luttinger Hamiltonian. Therefore, if $|2\gamma_2 k^2/m_0| >> |[\vec{k} \times \dot{\vec{k}}]/k^2|$, the general procedure described in the previous section is applicable and we may project Eq.(2) onto the rotating basis spanned respectively by the *HH* and *LH* spinor eigenfunctions of $H_{3/2}^{(M)}$. In the valence bands, the spin and orbital degrees of freedom are strongly coupled. Therefore, $\gamma_2 k^2/m_0$ may be considered as a small perturbation of Russell-Saunders-like states, *J* is approximately a good quantum number. Consequently, the projector onto the *LH* band is $P_{LH}^{(M)} = |J, 1/2><J, 1/2| + |J, -1/2><J, -1/2|$.

Hence, $P_{LH}^{(M)} J_{Z_M} P_{LH}^{(M)} = \sigma_{Z_M}/2$, $P_{LH}^{(M)} J_{X_M(Y_M)} P_{LH}^{(M)} = 1/2(J+1/2)\sigma_{X_M(Y_M)}$, which gives $\gamma_{LH,\|} = 1$, $\gamma_{LH,\perp} = J + 1/2 = 2$. Taking into account Eqs. (10), (11), and (13) we obtain

Taking into account Eqs. (10), (11), and (13) we obtain

$$\tilde{H}_{LH}^{(M)} = k^2/2m_L - \vec{\omega}^{(M)}\vec{\sigma}^{(M)} + \omega_{Z_M}\sigma_{Z_M}/2 = k^2/2m_L - q\vec{E}^{(M)}\vec{A}_{LH}^{(M)}, \quad (17a)$$

$$\vec{F}_{LH}^{(M)} = (3/2)(\hat{k}^{(M)}/k^2)\sigma_{Z_M}, \quad (17b)$$

where $m_L = m_0/(\gamma_1 + 2\gamma_2)$, $(\vec{A}_{LH}^{(M)})_\theta = \sigma_{Y_M}$, $(\vec{A}_{LH}^{(M)})_\varphi = [\cos\theta\,\sigma_{Z_M} - 2\sin\theta\,\sigma_{X_M}]/2$.

Similarly, for heavy holes $\gamma_{HH,\parallel} = 3$ and $\gamma_{HH,\perp} = 0$, therefore,

$$\tilde{H}_{HH}^{(M)} = k^2/2m_H - 3\omega_{Z_M}\sigma_{Z_M}/2, \qquad (18a)$$

$$\vec{F}_{HH}^{(M)} = -(3/2)(\hat{k}^{(M)}/k^2)\sigma_{Z_M}, \qquad (18b)$$

where $m_H = m_0/(\gamma_1 - 2\gamma_2)$, $(\vec{A}_{HH}^{(M)})_\theta = 0$, $(\vec{A}_{HH}^{(M)})_\varphi = 3\cos\theta\,\sigma_{Z_M}/2$. As expected, Eqs.(17) and (18) recover the results of Refs.[10 - 12]. The non-Abelian gauge structure is present only in the *LH* band. Moreover, in the Luttinger model the *HH* helicity is conserved, $[\hat{k}_{Z_M} S_{Z_M}, \tilde{H}_{HH}^{(M)}] = 0$, and is not affected by the adiabatic rotation of $\vec{k}$. This behavior is the consequence of the second order approximation made in the $\vec{k}\cdot\vec{p}$ perturbation theory that leads to the Luttinger Hamiltonian. Geometrically this result reflects the splitting of the Berry connection for states with the helicity difference $\Delta m > 1$, see Ref.[4] for details.

Calculation of the explicit form of $\Psi_C^{(M)}$ and, hence, $P_C^{(M)}$ and $\tilde{\gamma}_C^{(M)}$, as well as, an effective mass $m_C$ and g-tensor of an electron in the *C*-band is the straightforward theoretical problem[21]. It is instructive to compare $\Delta\tilde{\gamma}_C = \tilde{\gamma}_C - \vec{1}$ with $\Delta\vec{g}_C = \vec{g}_C - \vec{2}$. By definition $P_C^{(M)}(\vec{L}^{(M)} + 2\vec{S}^{(M)})P_C^{(M)} = \vec{g}_C^{(M)}\vec{\sigma}^{(M)}/2$, whereas $P_C^{(M)}(\vec{L}^{(M)} + \vec{S}^{(M)})P_C^{(M)} = \tilde{\gamma}_C^{(M)}\vec{\sigma}^{(M)}/2$. Therefore, *locally* in the k-space

$$(\Delta\vec{g}_C^{(M)} - \Delta\tilde{\gamma}_C^{(M)})\vec{\sigma}^{(M)}/2 = P_C^{(M)}\vec{S}^{(M)}P_C^{(M)} - \vec{\sigma}^{(M)}/2 \qquad (19)$$

and it is easy to see that in the *C*-band $\Delta\tilde{\gamma}_C$ and $\Delta\vec{g}_C$ differ only to the extent that the projection of the real spin $\vec{S}$ of an electron onto the Kramers-degenerate 2D space of states of the differs from the pseudo spin, see also Ref.[5]. This difference arises from the SO coupling and, hence, requires an admixture of the *p*-like valence band functions to $\Psi_C^{(M)}$ induced by the kinetic momentum. Consequently, if $\Delta_0 < E_g$, where $\Delta_0$ is the SO splitting of the valence bands, and the Fermi energy $E_F$ is much smaller than the fundamental energy gap $E_g$, i.e., for wide gap semiconductors and small electron concentrations, this difference is very small and can be neglected, $\Delta \vec{\gamma}_C \cong \Delta \vec{g}_C$. In III-V semiconductors, the main contribution to $\Psi_C^{(M)}$ comes from the upper valence bands of $\Gamma_8$ and $\Gamma_7$ symmetry. It is well known that in the "spherical" approximation at the bottom of the lowest *C*-band ($\Gamma_6$), the eight-band second order $\vec{k} \cdot \vec{p}$ perturbation theory gives[22]

$$\Delta g_C = -4m_0 P^2 \Delta_0 / 3E_g (E_g + \Delta_0), \qquad (20)$$

where $P = <S|p_X|P_X>/m_0$ is the Kane momentum matrix element describing the coupling of the *s*-like *C*-band and *p*-like valence band states. In this approximation, the gauge potential $\vec{A}_C^{(M)}$ is not pure, and there is covariant gauge field associated with it, $|\vec{F}_C| \sim |\Delta \gamma_\perp| = |\Delta g_C|$.

Consider, for example, the case of a constant uniform gate field applied to a bulk homogeneous semiconductor or a quasi-2D nanostructure. Without loss of generality, we may choose the direction of the field along the $z_L$-axis, $\vec{E} = (0, 0, E_{z_L})$. This field will drag a charge carrier moving along say $x_L$ out of the $x_L y_L$ plane, towards the surface and accordingly will rotate its wave vector. If the plane of $\vec{k}$ rotation remains constant, then the rotation axis can be assigned to $y_L = y_M$. In this case, substituting Eq.(1) into Eq.(10) we obtain

$$\tilde{H}_B^{(M)} = k^2/2m_B - \gamma_{\perp,B}\omega_{Y_L}\sigma_{Y_L}/2 = k^2/2m_B + \gamma_{\perp,B}(qE_{Z_L}k_{X_L}/k^2)\sigma_{Y_L}/2, \quad (21)$$

which for the *LH* gives the Eq.(7) of Ref.[11] and recovers the Eq.(25) of Ref.[12] for the *HH* and *LH*. Thus, with the obvious redefinition of the parameters, results obtained in these studies can be extended to the *C*-band. Furthermore, we can describe the effective spin Hamiltonian of the two-component Bloch spinor in the inertial frame. For *LH* and C bands the $H_B^{(L)}$ is given by

$$H_B^{(L)} = k^2/2m_B + \alpha_{B,SR}k_{X_L}\sigma_{Y_L}/2, \quad (22)$$

where we introduce $\alpha_{LH,SR} = eE_{Z_L}/k^2$ and $\alpha_{C,SR} = (-eE_{Z_L}/k^2)(\gamma_{C,\perp} - 1)$. Notably, the second term in Eqs.(19) and (20) is homologous to the Rashba Hamiltonian, which is fundamental for quasi-2D nanocrystals with structure induced asymmetry and may lead to a spin rotation and zero-field splitting in the *C* and *LH* bands[13,21]. Using $k = k_F = 10^8 m^{-1}$, $E = 10^6 V/m$, and $|\gamma_{\perp,C} - 1| \approx |\Delta g_{\perp,C}| \sim 10^{-1} \div 10^{-2}$ as typical values, we have $\alpha_{C,SR} = 10^{-11} \div 10^{-12} eV \cdot m$, which is comparable to the observed values in the lowest *C*-band associated with Rashba interaction under similar conditions (see, e.g., Ref.[21] and references therein).

**IV. Summary.**

Rotation of the crystal momentum of a charge carrier, induced by an external or a "built-in" electric field, violates the T-invariance of the system at $B = 0$ and leads to rotational distortion of the Bloch wave functions. In the rotating frame of reference, this deformation is represented by the action of Coriolis pseudo-force that is responsible for torque acting on the orbital momentum of a particle. In adiabatically isolated SO mixed Kramers-degenerate bands this perturbation gives rise to the specific form of spin-electric coupling - spin rotation - that may affect the coherent transport properties of a charge carrier and cause a spin precession in zero magnetic fields. Dynamic anisotropy of a system (*locally* in the $k$-space) is the fundamental precondition for materialization of these effects, which can be also interpreted in pure geometric terms as a manifestation of, in general, non-Abelian gauge potential in the momentum space. When SO coupling is suppressed, in the *C*-band of wide gap III-V semiconductors, the strength of the associated covariant gauge field is proportional to the effective electron $g$-tensor and is controllable by gate fields or a strain applied to the crystal.

---